\documentstyle[sprocl]{article}

\input{psfig}
\def\be{\begin{equation}}
\def\ee{\end{equation}}
\def\bea{\begin{eqnarray}}
\def\eea{\end{eqnarray}}

\begin{document}

\title{SUM RULES AND  2-QUARK FLUX-TUBE STRUCTURE}

\author{ A.M.GREEN, P.S.SPENCER }\address{Research Inst. for Theoretical 
Physics, University of Helsinki, Finland, 
green@phcu.helsinki.fi, spencer@rock.helsinki.fi }

\author{ C.MICHAEL }

\address{Theoretical Physics Division, Dept. of Math. Sciences, 
University of Liverpool, Liverpool, UK, cmi@liv.ac.uk}


\maketitle\abstracts{
Sum rules -- relating the static quark potential $V(R)$ 
to  the spatial distribution of the  action and energy in
the colour fields of  flux-tubes -- are applied in three ways: 
1) To extract  generalised $\beta$-functions:\\
2) As a consistency check for the use of excited gluon 
flux-tubes and as an estimate of the quark self-energies:\\
3) To extract approximate  sum rules using a simplified form of $V(R)$. \\
Also the flux-tube profiles are compared with hadronic string and  
flux-tube models.
                                }

\section{Introduction}
In ref.~\cite{cm} sum rules 
were derived to relate the sum over all spatial positions 
of colour fields (${\cal E, \ \cal B}$) to the static quark potential $V(R)$ 
and its derivative:   
\begin{equation}
{-1 \over b} \left( V+R {\partial V \over \partial R} \right) +
S_0= -\sum  ( {\cal E}_L + 2  {\cal E}_T  + 2  {\cal B}_T + {\cal B}_L) 
\label{TASU}
\end{equation} 
\begin{equation}
{1 \over 4 \beta f} \left( V+R {\partial V \over \partial R} \right) + 
E_0 = \sum  ( - {\cal E}_L + {\cal B}_L) 
\label{TELSU}
\end{equation} 
\begin{equation}
{1 \over 4 \beta f} \left( V-R {\partial V \over \partial R} \right) +
E_0 = \sum  ( - {\cal E}_T  +   {\cal B}_T ) \ .
\label{TEPSU}
\end{equation} 
Here $L(T)$ refers to longitudinal(transverse) with respect to
the interquark separation axis. The  $ {\cal E}({\cal B})$ have the 
interpretation of  gauge
invariant averages of the fluctuation of squared  strengths of the
colour electric(magnetic)fields. The parameters $b$ and $f$ are related to 
the  generalised  $\beta$-functions 
 \begin{equation}
{\partial \beta_{ij} \over \partial \ln a_k}=
S \ \hbox{if}\ k=i \ \hbox{or}\ j
\ \ \hbox{and} \ \
{\partial \beta_{ij} \over \partial \ln a_k}=
U \ \hbox{if}\ k\ne i \ \hbox{or}\ j
\end{equation}                                         
by the expressions
\begin{equation}
 b=2(S+U)=-0.3715(1+0.49 \alpha+\dots ) 
\ \ \ , \ \ \ \  2\beta f = U-S = 2 \beta(1-1.13\alpha+ \dots) \ ,
\end{equation}
where the perturbative series for these quantities 
in terms of the bare lattice coupling $\alpha=g^2/4\pi=1/\pi\beta$ 
for $SU(2)$ colour fields  are also given~\cite{karsch}.
The $S_0$ and $E_0$ are the self energy terms.
The general strategy for utilizing these sum rules is to insert known
values or forms of $V(R)$ on the LHS and to measure on a lattice the 
${\cal E, \ B}$ on the RHS. In this talk three such applications are
discussed. In addition to these integrated colour field quantities, the
individual field profiles are also studied.

\section*{ 2 Three applications of the above sum rules}
The $V(R)$, ${\cal E}$ and ${\cal B}$ are all calculated in quenched SU(2)
on the same $16^3\times 32$ lattice with $\beta=2.4$ using basis states with
different degrees of fuzzing. The latter enables the excited gluon states 
with cubic symmetry $E_u$ and $A_{1g}'$ to be studied.
\subsection{$V(R)$ directly from the lattice to give Beta-functions}
With $V(R)$ calculated on the same lattices as the  ${\cal E,  B}$ ,
combinations of the above sum rules are made at two values of $R$ to eliminate 
the ${\partial V \over \partial R}$ and also the self-energy terms giving: 
\be
f=-{\left[V(R_1)-V(R_2)\right]
\over
2 \beta  \left[ (A+B)_{R_1}- (A+B)_{R_2} \right]}
\ee
\be
b 
={ 2 [ V(R_1)-V(R_2)] \left[ 1+{ 
 A_{R_1}- A_{R_2}
 \over 
 B_{R_1}- B_{R_2}}
\right]^{-1}
\over
C_{R_1}-C_{R_2}
} \ ,
\label{AB}
\ee
where
\[A_R=\sum  ({\cal E}_T  -   {\cal B}_T )_R  \ \ , \ \
B_R=\sum ({\cal E}_L - {\cal B}_L)_R              \]
\[C_R= \sum  ( {\cal E}_L + 2  {\cal E}_T  + 2  {\cal B}_T + {\cal B}_L)_R.\]
This results in best estimates of $b=$ --0.35(2) and $f=$ 0.61(3). Therefore,
$b$ is seen to be far from the perturbation estimate of --0.42 with
$\alpha_{effective}=0.26$.
A more detailed account can be found in ref.~\cite{letter}
\subsection{A parametrization of the lattice potential $V(R)$}
Armed with estimates for $b$ and $f$, Eqs.~\ref{TASU}-\ref{TEPSU} can be
studied separately using the following interpolations of the lattice $V(R)$ 
for the ground ($A_{1g}$) and first excited ($E_u$) state:
 \begin{equation}
 V(R)_{A1g}=0.562 + 0.0696 R - 0.255/R -0.045/R^2 
 \end{equation}
 \begin{equation}
  V(R)_{Eu} - V(R)_{A1g} = \pi/R -4.24/R^2 + 3.983/R^4.
 \end{equation}
For the ground state case it is found that all three sum rules are well
satisfied with $S_0$=0 and $E_0$=0.1. However, the comparison is not so good
for the $E_u$ state -- indicating that this state is still somewhat contaminated
for the values of Euclidean time $T \le 4$, for which a signal could be
measured.  More details can be found in ref.~\cite{GMP}
\subsection{A simple parametrization of $V(R)$}

If the form $V(R)=c+b_sR+e/R$ is used in the sum rules, then they reduce to
\be
\label{s}
c+ 2b_sR=\sum 2(S+U)[{\cal E}_L+2{\cal E}_P+2{\cal B}_P+{\cal B}_L] 
\ee
\be
\label{el}
c+2b_sR=\sum 2(S-U)[{\cal E}_L-{\cal B}_L] \ {\rm and} \  
c+\frac{2e}{R}=\sum 2(S-U)[{\cal E}_P-{\cal B}_P], 
\ee
where $e\approx -0.25 (\pi)$ for the ground($E_u$) state.
Since Eqs.~\ref{s} and \ref{el}a are independent of $e$, they reduce further
to
\be
\label{sr5}
\sum_{x,y,z} [S^{\rm Ex}]= \sum_{x,y,z} [S^{\rm GS}] \quad{\rm and}\quad
\sum_{x,y,z} [E_L^{\rm Ex}]= \sum_{x,y,z} [E_L^{\rm GS}] 
\ee
and the third becomes
\be
\label{sr6}
\frac{\Delta e }{R}\approx \beta f \sum_{x,y,z} [E^{\rm Ex}-E^{\rm
GS}],
\ee
where $S^i(E^i)$ refers to the action(energy) combinations of
${\cal E}({\cal B})$ in Eqs.~~\ref{s}, \ref{el}.
For sufficiently large $R$, the replacement
$\sum_{x,y,z} \rightarrow \frac{R}{a}\sum_{x,y}$ should be a reasonable
approximation. In this case the 3-d sum rules in Eqs.~\ref{sr5},\ref{sr6}
reduce to 2-d sum rules involving less Monte Carlo data and are useful
when discussing flux-tube profiles in the next section.
 More details can be found in ref.~\cite{GMP}

\section*{3  Flux-tube profiles}
So far only the sum rules for combinations of ${\cal E, \ B}$  
have been discussed. In this section the non-integrated combinations
of  ${\cal E,  B}$ are compared  with two models.  
In Fig. 1a
 comparison is made for energy profiles at the centre of the line connecting 
the two quarks. There MC refers to the Monte Carlo result at $R=6$, 
 BBZ to the dual superconducting model of ref.~\cite{BBZ} and  
 IP to the string motivated model of ref~\cite{IP}. In the latter, 
the string energy was
tuned to fit the MC result on the central axis. The disagreement with the
other two curves is not surprising, since the IP model includes a sizeable
zero-point energy in its description of the gluon fields. Therefore, only
{\em differences} between the profiles of the separate states should be
compared. The similarity between the MC and BBZ results can then be interpreted
as the  two having similar self-energies .
In Figs.~1b the
action profiles for the ground and $E_u, A_{1g}'$ states are shown. Here, it
is of interest to see that the $A_{1g}'$ case has a dip at $r\approx 2a$ and 
is reminiscent of the node expected in excited s-wave states. The total action 
and energy profiles for the ground state are in agreement with 
refs.~\cite{bali} \ \cite{hay}.

The authors wish to  acknowledge that these calculations were carried
out at the Centre for Scientific Computing's C94 in Helsinki and the
RAL(UK) CRAY Y-MP and J90.  This work is part of the EC Programme
``Human Capital and Mobility'' -- project number ERB-CHRX-CT92-0051.

\begin{figure}
\includegraphics{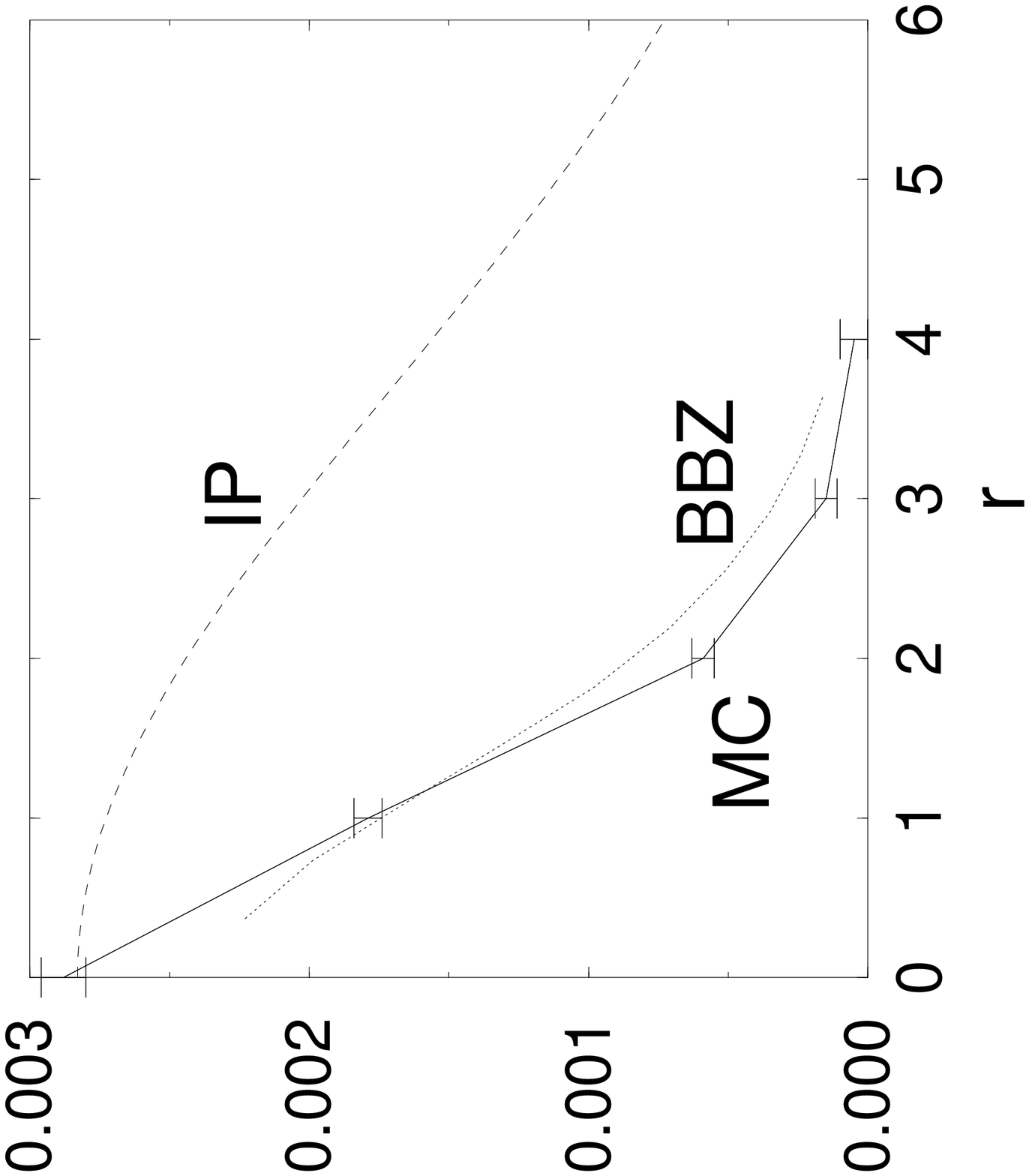}
\hspace{6.8cm}\includegraphics{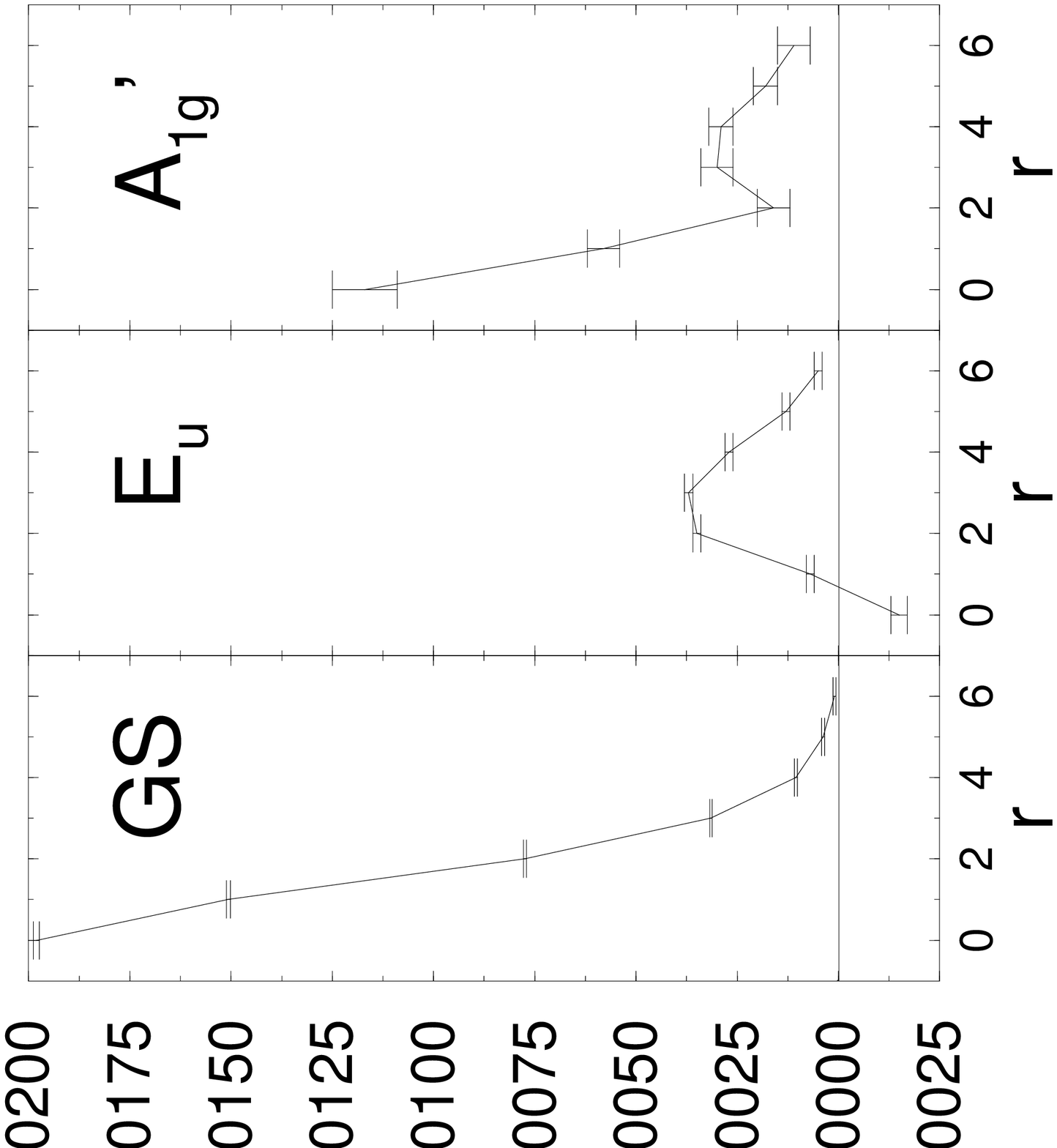}
\vspace{3.5cm}\caption{(a) The Energy profile $E(r)$ for $R=6$:
(b) The Action profiles $S(r)$ for 
the states A$_{1g}$, E$_u$ and A$_{1g}'$, calculated at $R=6$. 
Both $E(r)$ and $S(r)$ are dimensionless and $r$ -- the distance from the
central axis -- is in
lattice units of $a=0.12$fm.
\label{fig:flux}}
\end{figure}



\section*{References}
\begin{thebibliography}{99}
\bibitem{cm} C. Michael,  {\em Phys.Rev.}   {\bf D53}, 4102  (1996).

\bibitem{karsch} F. Karsch, {\em Nucl. Phys.}  {\bf B205}, 285  (1982) .

\bibitem{letter} C. Michael, A.M. Green and P.S. Spencer, hep-lat/9606002
-- to be published in {\em Phys.Lett.B} 
\bibitem{GMP} A.M. Green,  C. Michael and P.S. Spencer, in preparation

\bibitem{BBZ} M. Baker, these Proceedings; M. Baker, J.S. Ball and F. Zachariasen,
{\em Int.Jour.Mod.Phys.}   {\bf A11}, 343 (1996)

\bibitem{IP} N. Isgur and J.E. Paton, {\em Phys. Rev.}   {\bf A31}, 2910  (1985).

\bibitem{bali} G. Bali, K. Schilling and C. Schlichter, 
{\em Phys. Rev.}  {\bf D51}, 5165  (1995).

\bibitem{hay} R.W.~Haymaker, V.~Singh and Y. Peng, 
  {\em Phys.Rev.}  {\bf D53}, 389  (1996).
  

\end{thebibliography}
\end{document}